# Applicability and non-applicability of equilibrium statistical mechanics to non-thermal damage phenomena: II. Spinodal behavior

Short title: Applicability II…

S.G. Abaimov[1]

[1]Pennsylvania State University, University Park, PA, 16802, USA

E-mail: sgabaimov@gmail.com and sua20@psu.edu

**Abstract:** This paper investigates the spinodal behavior of non-thermal damage phenomena. As an example, a non-thermal fiber-bundle model with the global uniform (meanfield) load sharing is considered. In the vicinity of the spinodal point the power-law scaling behavior is found. For the meanfield fiber-bundle model the spinodal exponents are found to have typical meanfield values.

## 1 Introduction

Damage as a complex phenomenon has been studied by many authors and a survey of recent developments in damage mechanics can be found in [1-3]. Many attempts [4-12] have been made to apply equilibrium statistical mechanics to damage phenomena. However, damage phenomena usually exhibit more complex behavior than gas-liquid or magnetic systems. The reason is that damage has two different forms of appearance.

Firstly, damage behavior inherits thermal fluctuations from the media in which it occurs. The main representative of thermal damage fluctuations is the Griffith theory. The application of statistical mechanics here is straightforward and has many parallels with gas-liquid systems. Secondly, even in the case of non-thermal systems the occurrence of damage has complex topological appearance. This behavior can also be described by the formalism of statistical mechanics [12]. However, all resulting equations in this case are valid not for energy characteristics of damage but for its topological properties. This type of behavior is often observed when the dynamical time scale of fracture is much faster than the time scale of thermal fluctuations and conductivity. In this case *a priori* input disorder in a model plays the crucial role.

In many systems these two different forms of damage appearance usually occur simultaneously and make it difficult to investigate possible system behavior. Therefore, it would be reasonable to separate these two different types of behavior and study them one by one. Later, combined, they can provide extensive explanations for a wide range of physical phenomena. Therefore, in this paper we will follow the approach developed earlier in [12] and will study only the topological behavior of non-thermal systems with damage.

As it was suggested by many authors, the point of singularity $\frac{d\sigma}{d\varepsilon}=0$ is a spinodal point S (Fig. 1) of a system [4, 6, 12, 13]. It was found that thermodynamic systems near their spinodals exhibit power-law dependences similar to the critical [14]. However, there are differences between critical and spinodal behavior. In the vicinity of critical point temperature, as a field parameter, plays the crucial role and many scaling laws depend on the difference of temperature from the critical. In the vicinity of spinodal the crucial role is played by another field parameter – magnetic field or pressure. Therefore thermodynamic systems in the vicinity of the spinodal exhibit power-law dependences on the difference of magnetic field from its spinodal value.

In the case of systems with damage we have a geometric constraint that the total force $\sigma$ on the model surface depends directly on the strain $\varepsilon$ and the total amount of intact material $(1 - D)$. For the fiber-bundle model this dependence will be illustrated by Eq. (1) below. As it was demonstrated by Abaimov [12], this constraint is so powerful that it even acts on the level of microstates. The presence of this constraint indicates that in comparison with the thermodynamic systems the non-thermal topological damage behavior has one field parameter less. In fact, the constraint leaves only one field parameter – the total force $\sigma$ acting on the model surface. Therefore it is expected that in the vicinity of the spinodal point $\frac{d\sigma}{d\varepsilon} = 0$ the systems with damage exhibit scaling power-law behavior of different quantities relative to the difference of external force $\sigma$ from its spinodal value: $\Delta\sigma = \sigma - \sigma_S$.

This paper develops a rigorous, systematic approach to investigate the scaling behavior of damage phenomena in the vicinity of the spinodal point. In section 2 we introduce a model which we will use to illustrate these concepts. Also we define the concepts of a microstate, macrostate, and equilibrium state. In section 3 we investigate the spinodal behavior of an order parameter. In section 4 we obtain dynamical properties of the model and find critical-spinodal slowing-down. Section 5 demonstrates which quantity should be utilized as a susceptibility and investigates its scaling properties. Also in this section we look at the analogue of the specific heat and its spinodal exponent. In section 6 we discuss how we can find a free energy potential of the metastable state. For the introduced potential scaling properties are investigated.

**2 Model**

Damage is a complex phenomenon. It can be associated with local and non-local load sharing, brittle and ductile behavior. It can emerge both in one-dimensional and three-dimensional systems, leading in the last case to three-dimensional stress patterns of crack formation. The basic principles of damage are often completely disguised by the secondary side effects of its appearance.

Therefore, to investigate the spinodal behavior, it is reasonable to consider initially a simple model. So, in statistical mechanics the Van der Waals' meanfield model is usually used to illustrate the behavior of gas-liquid systems. In the case of magnetic systems an analogy is the meanfield Ising model with the infinite range of interactions. The basic principles of scaling behavior can be illustrated by these simple models. Further models, corresponding to real systems, can be constructed as more accurate and more complex improvements.

In this paper we will investigate the meanfield approximation of damage phenomena. As a model we consider a static (deterministic, quenched) fiber-bundle model (further FBM) with the uniform global (meanfield) load sharing. The term 'static' as an opposite to the term 'stochastic' is used to specify that each fiber has *a priori* assigned strength threshold $s$ which does not change during the model evolution, and this fiber can fail only when its stress $\sigma_f$ exceeds its strength $s$. All fibers have predefined strengths, distributed with *a priori* specified probability density function $p_s(s)$. The cumulative distribution function $P_s(\sigma_f) = \int_0^{\sigma_f} p_s(s)ds$ is the probability for a fiber to have been already broken if its stress is supposed to be $\sigma_f$.

We assume that there is an ensemble of identical systems. The systems are non-thermodynamic and thermal fluctuations are naturally absent in them. Instead, each system in the ensemble realizes some strength distribution over its fibers. Each system in the ensemble fails in its own way. This particular system does not exhibit any variability because the way it fails is prescribed to it in advance and forever. However, different systems in the ensemble fail in different ways. This introduces stochastic topological fluctuations. The model is deterministic (static) in the sense that for each particular system the realization of strength distribution is assigned *a priori* and does not change during the system evolution. Therefore for the same realization of strength distribution the system follows during its evolution exactly the same deterministic trajectory. However, the model is stochastic in the sense that *a priori* assigned sample distribution of a system is prescribed stochastically and only on the ensemble average should correspond to $P_s(s)$.

We assume that the number of fibers in the model $N$ is constant and infinite in the thermodynamic limit. Intact fibers carry all the same strain $\varepsilon_f$ which is identically equal to the strain $\varepsilon$ of the total model as a 'black box': $\varepsilon_f \equiv \varepsilon$. The stress of each intact fiber is assumed to have the linear elastic dependence on the strain till the fiber failure: $\sigma_f = E\varepsilon_f$. The Young modulus $E$ here is assumed to be the same for all fibers. This introduces the concept of non-linear stress-strain dependence for the total model although each fiber behaves elastically till its failure. We introduce the damage parameter $D$ as the fraction of broken fibers. If the total number of fibers in the model is $N$ then the number of broken fibers is $ND$ and the number of intact fibers is $N(1-D)$. If we, as an external observer, were to look at the model as at a 'black box' we would not know how many fibers are broken inside and we would see that the external force is applied to $N$ fibers, creating uniform 'virtual' stress $\sigma$ at the model surface. The actual stress in fibers is $1/(1-D)$ times higher due to the fiber failure and stress redistribution. Therefore, the virtual stress of the model (how an external observer sees it) is $\sigma = (1 - D)\sigma_f$ or

$$\sigma = (1 - D)E\varepsilon. \tag{1}$$

It is important to note here that this 'geometric' constraint is so strong that it acts not only on the level of equilibrium quantities but even on the level of each microstate (each realization of strength distribution). As we will see later, this significantly changes the behavior of the model in contrast to gas-liquid systems. One of the direct consequences is that both strain $\varepsilon$ or stress $\sigma$ could be chosen to be a field parameter. However, in contrast to other systems, only one of these parameters is independent. 'Geometric' constraint (1) and an equation of state make two of three parameters $D$, $\varepsilon$, $\sigma$ dependent.

In the case of gas-liquid systems microstates are defined as cells in the phase space of a system. In the case of magnetic systems microstates are defined as different realizations of "up" and "down" spins on a lattice: ↑↑↓↓↑. We can define microstates for the FBM in an analogous way as different realizations of broken and intact fibers.

So, for the FBM with $N = 3$ fibers all possible microstates are |||, ||x, |x|, x||, |xx, x|x, xx|, and xxx where symbol '|' denotes an intact fiber while symbol 'x' – a broken fiber. Further in the paper index $n$ will be used to enumerate all possible microstates $\{n\}$.

For the specified damage $D$ and for the external boundary constraints (further BC) $\varepsilon$ = const or $\sigma$ = const each microstate has the probability

$$w_{\{n\}}^{equil} \equiv w^{equil}(D) = (1 - P_s(E\varepsilon))^{N(1-D)} (P_s(E\varepsilon))^{ND} \quad \text{for } \varepsilon = \text{const or} \tag{2a}$$

$$w_{\{n\}}^{equil} \equiv w^{equil}(D) = \left(1 - P_s\left(\frac{\sigma}{1-D}\right)\right)^{N(1-D)} \left(P_s\left(\frac{\sigma}{1-D}\right)\right)^{ND} \quad \text{for } \sigma = \text{const} \tag{2b}$$

as the probability that $N(1 - D)$ fibers are intact and $ND$ fibers are broken. This probability $w_{\{n\}}^{equil}$ is dictated by the prescribed BC $P_s(s)$. This BC is a model input and acts similar to the temperature prescribed in the canonical ensemble. An external medium dictates the equilibrium distribution of probabilities but the system actually can realize itself in a non-equilibrium state with any other probability distribution $w_{\{n\}}$. Only the equilibrium state is dictated by the BC $P_s(s)$ therefore we used abbreviation '*equil*' to emphasize that this probability distribution corresponds to the equilibrium with the BC $P_s(s)$.

In general, a macrostate may be defined as a union of all possible microstates realized with the specified probabilities. However, in this paper a simpler definition will be used when these probabilities equal only to zero or unity. In other words, further in this paper as to a macrostate we will refer to a subset of microstates chosen by a particular property. For example, the definition of a macrostate in the thermal canonical ensemble is a subset of all microstates with the specified energy. For the FBM we will use the definition of a macrostate [D] as a subset of all microstates with the specified fraction of broken fibers, *i.e.*, with the specified damage D: $[D] = \bigcup_{\{n\}: D_n = D} \{n\}$.

All microstates corresponding to the macrostate [D] have the same probability (2) and the number of these microstates is given by the combinatorial choice of ND broken fibers among N fibers

$$g_{[D]} = \frac{N!}{(ND)!(N(1-D))!} \approx_{\ln} \frac{1}{D^{ND}(1-D)^{N(1-D)}} \tag{3}$$

where symbol "$\approx_{\ln}$" means that in the thermodynamic limit $N \to +\infty$ all power-law multipliers are neglected in comparison with the exponential dependence on N. Everywhere further symbol "$\approx_{\ln}$" will mean the accuracy of the exponential dependence neglecting all power-law dependences. For the logarithm of such equations we will use symbol "$\approx$".

To find the total probability of the macrostate D we need to multiply the probability of each microstate (2) by the total number of microstates (3) corresponding to this macrostate

$$W_{[D]}^{equil}(D) = \sum_{n=1}^{g_{[D]}} w_{\{n\}}^{equil}(D) = g_{[D]} w^{equil}(D) \approx_{\ln} \left(\frac{1 - P_s(E\varepsilon)}{1 - D}\right)^{N(1-D)} \left(\frac{P_s(E\varepsilon)}{D}\right)^{ND} \tag{4a}$$

for the BC $\varepsilon$ = const and

$$W_{[D]}^{equil}(D) \approx_{\ln} \left(\frac{1 - P_s\left(\frac{\sigma}{1-D}\right)}{1-D}\right)^{N(1-D)} \left(\frac{P_s\left(\frac{\sigma}{1-D}\right)}{D}\right)^{ND} \tag{4b}$$

for the BC $\sigma$ = const. This is the probability for the macrostate $D$ to be observed in equilibrium with the BC $P_s(s)$.

We used above the term 'equilibrium' but did not specify what we refer to by this term. The wrong way would be to imagine a time-dependent system in some detailed balance with the BC $P_s(s)$. There is no time dependency in the model. Each system fails only once in *a priori* prescribed way. Actually, this particular system is only one microstate. By the term 'equilibrium' we refer to the ensemble of failed systems whose stochastic properties on the average correspond to the prescribed BC $P_s(s)$. For the equilibrium we will use two different definitions. The BC $P_s(s)$ is assumed to prescribe the equilibrium probability distribution $w_{\{n\}}^{equil}$ for all microstates. Therefore, the equilibrium with this BC could be identified with a system which can realize itself on all microstates with equilibrium probabilities: $w_{\{n\}} = w_{\{n\}}^{equil}$. In other words, all microstates are possible but their probabilities are dictated by the BC $P_s(s)$. As of an example we could think of the equilibrium in the thermal canonical ensemble where all microstates with all energies are possible but their probabilities are dictated by Boltzmann distribution. The superscript '*equil*' will be used for this definition. Then the value of any quantity $f$ in equilibrium with the BC $P_s(s)$ by definition is

$$\langle f \rangle^{equil} \equiv \sum_{\{n\}} w_{\{n\}}^{equil} f_{\{n\}} .$$

In contrast, another definition is the equilibrium (most probable) macrostate, *i.e.*, a system that can realize itself only on (that is isolated on) a subset of microstates corresponding to the most probable macrostate. This is the macrostate which gives the main contribution to the partition function. As an example we could think of the equilibrium macrostate of the thermal canonical ensemble where we count only those microstates whose energies equal to the equilibrium value $E_0 = Nk_BT/2$. To distinguish this case the subscript '$_0$' will be used.

As an example, we can consider the equilibrium value of the damage parameter. As to $\langle D \rangle^{equil}$ we refer to the damage parameter averaged over the equilibrium distribution of probabilities $\langle D \rangle^{equil} = \sum_{\{n\}} D_{\{n\}} w^{equil}_{\{n\}} = \sum_{[D]} g_{[D]} D w^{equil}(D)$. As to $D_0$ we refer to the damage corresponding to the most probable microstate: $W^{equil}_{[D_0]}(D_0) = \max_D W^{equil}_{[D]}(D)$. Of course, in the thermodynamic limit these quantities are equal.

## 3 Scaling of the order parameter

We return now to the investigation of spinodal point $\frac{d\sigma}{d\varepsilon} = 0$. First we need to define an order parameter. Order parameters in gas-liquid systems are densities of phases; in magnetic systems they are magnetizations of phases. Similar to this approach we define different phases of damage as phases with different fractions of broken fibers. The damage parameter $D$ will play the role of the order parameter distinguishing phases.

If we assume for the general case the absence of a singularity of $P_s(s)$ at the spinodal point S then in the vicinity of the spinodal point the stress-damage dependence can be approximated by a parabolic descent: $\sigma - \sigma_S = - C(D_0 - D_{0S})^2$ where $C$ is some positive constant. This immediately gives that the spinodal exponent $\beta$ of the order parameter $D$ has the typical meanfield value 1/2 [13, 15]:

$$\left|D_0 - D_{0S}\right| \propto \left|\sigma - \sigma_S\right|^\beta \propto \sqrt{\left|\sigma - \sigma_S\right|}. \tag{5}$$

To derive this formula we used only general assumptions about the equilibrium dependence and did not refer to any particular boundary conditions. Therefore this scaling is valid for both external BCs $\varepsilon$ = const and $\sigma$ = const. Remembering Eq. (1), we, as it was suggested by many authors, obtain the same parabolic approximation for the stress-strain dependence: $\sigma - \sigma_S \propto -(\varepsilon - \varepsilon_S)^2$.

**4 Dynamical slowing-down**

We will follow in this section Pradhan *et al.* [15] and Bhattacharyya *et al.* [16]. However, in contrast to these references, which considered particular distributions $P_s(s)$, we consider the general case of an arbitrary dependence $P_s(s)$.

First we consider the BC $\sigma$ = const. If at a time step $t$ the fraction of intact fibers is $(1 - D_t)$ then at this time step the stress of each fiber is $\sigma_{f,t} = \sigma / (1 - D_t)$. We assume the model to be discretely inductive. Therefore at the next time step the fraction of broken fibers is $D_{t+1} = P_s(\sigma_{f,t})$. For the iterative equation for $D_t$ we obtain $D_{t+1} = P_s(\sigma / (1 - D_t))$.

Fluctuations around the equilibrium state are proportional to $1/\sqrt{N}$ far from spinodal and to $1/\sqrt[4]{N}$ in the vicinity of spinodal. In both cases the averaged size of fluctuations is zero in the thermodynamic limit. Therefore, the avalanches, caused by deviations from $P_s(s)$, have non-zero probabilities to occur only if their sizes are small. Far from the spinodal point at the metastable state A (Fig. 1) we can use linear approximation for the function $P_s(s)$. Therefore, $D_{t+1} - D_A = P_s'|_A \frac{\sigma_A}{(1-D_A)^2}(D_t - D_A)$ where $P_s'|_A$ is the derivative of function $P_s(x)$ taken at the metastable state $D_0 = D_A$. For the equilibrium we have $D_0 = P_s(\sigma/(1-D_0))$. The differential of this equation is

$$dD_0 = P_s'\left\{\frac{d\sigma}{1-D_0} + \frac{\sigma dD_0}{(1-D_0)^2}\right\} \text{ or } P_s' = \left\{\frac{1}{1-D_0}\frac{d\sigma}{dD_0} + \frac{\sigma}{(1-D_0)^2}\right\}^{-1}. \tag{6}$$

Therefore for the iterational equation of $\Delta D_t \equiv D_t - D_A$ we obtain $\Delta D_{t+1} = \Delta D_t / \left\{1 + \frac{1-D_A}{\sigma_A}\frac{d\sigma}{dD_0}\Big|_A\right\}$ where $\frac{d\sigma}{dD_0}$ is the change of the external force $\sigma$ with the change of the equilibrium damage $D_0$. Far from the spinodal point (assuming monotonic growth of $\sigma$ with the growth of $D_0$ before the spinodal) the derivative $\frac{d\sigma}{dD_0}$ is positive and does not equal zero. Therefore indeed the linear approximation is enough and for the iterational equation we obtain

$$\Delta D_{t+1} - \Delta D_t = \Delta D_t \left(\left\{1 + \frac{1-D_A}{\sigma_A}\frac{d\sigma}{dD_0}\Big|_A\right\}^{-1} - 1\right) = -\frac{\Delta D_t}{1 + \frac{\sigma_A}{1-D_A}\frac{dD_0}{d\sigma}\Big|_A} \text{ or } \frac{d\Delta D}{\Delta D} = -\frac{dt}{1 + \frac{\sigma_A}{1-D_A}\frac{dD_0}{d\sigma}\Big|_A}.$$

This equation has the solution of the exponential decay $\Delta D \propto \exp\{-t/t_{ref}\}$ where $t_{ref} = 1 + \frac{\sigma_A}{1-D_A}\frac{dD_0}{d\sigma}\Big|_A$. Therefore far from the spinodal point the attenuation of avalanches is exponential: after the break of the initial fibers the number of breaking fibers decay exponentially with time.

In contrast, in the vicinity of the spinodal point S the system exhibits a significantly different type of behavior. At the spinodal point $\frac{d\sigma}{dD_0}\Big|_S = 0$ or $t_{ref} = +\infty$.

This is the critical-spinodal slowing-down: the relaxation time becomes infinite. Let us be non-rigorous for a second. Assuming that $P_s(s)$ does not have a singularity at the spinodal point, for the derivative $\frac{d\sigma}{dD_0}$ in the vicinity of the spinodal point we may use the linear approximation $\left.\frac{d\sigma}{dD_0}\right|_A \approx \left.\frac{d\sigma}{dD_0}\right|_S + \left.\left(\frac{d}{dD_0}\frac{d\sigma}{dD_0}\right)\right|_S (D_A - D_S) = \left.\frac{d^2\sigma}{dD_0^2}\right|_S (D_A - D_S)$.

Therefore $t_{ref} \propto |D_A - D_S|^{-1} \propto |\sigma_A - \sigma_S|^{-1/2}$.

However, in the analysis above we were not rigorous because at the spinodal point the linear approximation $D_{t+1} - D_A = \left.P_s'\right|_A \frac{\sigma_A}{(1-D_A)^2}(D_t - D_A)$, used above, is already not enough and we have to use the quadratic approximation

$D_{t+1} - D_S = \left.P_s'\right|_S \frac{\sigma_S}{(1-D_S)^2}(D_t - D_S) + \frac{(D_t - D_S)^2}{2}\left\{\left.P_s''\right|_S \frac{\sigma_S^2}{(1-D_S)^4} + \left.P_s'\right|_S \frac{2\sigma_S}{(1-D_S)^3}\right\}$. Remembering

Eq. (6) and that $\left.\frac{d\sigma}{dD_0}\right|_S = 0$ we obtain $\left.P_s'\right|_S = \frac{(1-D_S)^2}{\sigma_S}$. Differentiating (6) second time we obtain $\left.P_s''\right|_S = -\left(\frac{1-D_S}{\sigma_S}\right)^3 \left\{\left.\frac{d^2\sigma}{dD_0^2}\right|_S + \frac{2\sigma_S}{(1-D_S)^2}\right\}$. To find $\left.\frac{d^2\sigma}{dD_0^2}\right|_S$ we need to differentiate

Eq. (1) two times to obtain $\left.\frac{d^2\sigma}{dD_0^2}\right|_S = \left(\frac{\sigma_S}{(1-D_S)^2}\right)^2 \left.\frac{d^2\sigma}{d(E\varepsilon)^2}\right|_S$. For the iterational equation of

$\Delta D_t \equiv D_t - D_S$ we obtain $\Delta D_{t+1} - \Delta D_t = \frac{\Delta D_t^2}{2}\left\{-\left.\frac{d^2\sigma}{d(E\varepsilon)^2}\right|_S \frac{\sigma_S}{(1-D_S)^5} - \frac{2}{(1-D_S)^3} + \frac{2}{1-D_S}\right\}$ or

$\frac{d\Delta D}{\Delta D^2} = \frac{dt}{2}\left\{-\left.\frac{d^2\sigma}{d(E\varepsilon)^2}\right|_S \frac{\sigma_S}{(1-D_S)^5} - \frac{2}{(1-D_S)^3} + \frac{2}{1-D_S}\right\}$. The solution of this equation is

$$\Delta D = -\cfrac{1}{const_1 + \cfrac{t}{2}\left\{-\left.\cfrac{d^2\sigma}{d(E\varepsilon)^2}\right|_S \cfrac{\sigma_S}{(1-D_S)^5} - \cfrac{2}{(1-D_S)^3} + \cfrac{2}{1-D_S}\right\}} = -\cfrac{1}{const_1 + const_2 t}. \qquad (7)$$

This dependence is known as Omori law in earthquake statistics. If we associate each particular fiber failure with an earthquake, this dependence shows the time dependence of the number of aftershocks after the mainshock (the initial fiber failure, which has started the avalanche). We see that this dependence appear only in the vicinity of the spinodal point. This supports the point of view in earthquake statistics that the behavior of earthquakes is described by a system in the vicinity of its spinodal or critical point.

In the limit of infinite time the dependence of Eq. (7) becomes $\Delta D \propto -\cfrac{1}{const_2 t}$. This gives the spinodal exponent $\tau = 1$ for the scaling $|\Delta D| \propto \cfrac{1}{t^\tau}$. This is the slowing-down – a well known phenomenon in the physics of critical behavior. In the vicinity of the critical point all processes slow down with the infinite time for final relaxation. In other words, the characteristic time of the exponential decay becomes infinite and the exponential decay is changed to the power-law relaxation.

Because of this similarity with the critical phenomena many authors [15-19] supported the point of view that the point S is not a spinodal point but a critical point. As we have shown in [12], the phenomenon of damage is indeed very strange. From one point of view the state A in Fig. 1 is metastable: a crack with the critical size causes the nucleation process. The state B is unstable: the derivative $\left.\cfrac{d\sigma}{dD_0}\right|_B$ is negative. Therefore the curve A – S – B is similar to the curve metastable-spinodal-unstable for the meanfield homogeneous gas-liquid or magnetic systems. From another point of view the typical spinodal behavior in statistical mechanics is when one local minimum of the free energy potential with the change of the field parameter becomes more and more shallow and finally disappears. As we illustrated in [12], the behavior of damage is significantly different: the local minimum does not become shallower and does not disappear. Contrary, it keeps its depth unchanged and at the point S coalesces with the unstable minimum which has the same depth. This is more typical for the critical phenomena. Therefore, we see that the phenomenon of damage does not exhibit the presence of critical or spinodal points in the strict sense of statistical mechanics. In contrast, it exhibits the presence of a new type of the spinodal-critical point with its own intrinsic properties.

For the BC $\varepsilon$ = const there is no dynamical behavior. For each prescribed value of the external strain $\varepsilon$ the value of the damage is determined *a priori*. The slow increase of strain (if $P_s(s)$ is a smooth function) causes fibers to fail one by one without avalanches. Therefore the size of all avalanches is unity and the spinodal exponent is $\tau = 0$.

**5 Susceptibility, specific heat, and fluctuation-dissipation theorem**

Susceptibility appears to be one of the most important quantities for damage phenomena [20]. Any metastable state is able to cause the global fracture in the presence of a crack of critical size. However, only the spinodal point S usually represents the actual, catastrophic danger because the size of the critical crack vanishes at this point. Therefore it is of primary necessity to know how far a system is from the point of spinodal. And the susceptibility plays a key role in this analysis. In accordance with the fluctuation-dissipation theorem the susceptibility has two different forms of scientific meaning. Firstly, it is proportional to the integral of a correlation function and to the variance of fluctuations. Both these quantities diverge (stop to be Gaussian) in the vicinity of the spinodal point which causes the divergence of the susceptibility. Secondly, the susceptibility represents the response of system's order parameter to small perturbations of its field parameter. If we expect this response to diverge in the vicinity of the global fracture, this gives us a method of prediction of catastrophes. Indeed, causing small, non-destructive perturbations of the field parameter and watching the system response we may be able to predict the reliability of the system.

We first assume that the system is under the most interesting BC $\sigma$ = const. For magnetic systems the susceptibility determines how fast the averaged magnetization of the system changes with the change of the external magnetic field: $\chi \propto \frac{\partial m}{\partial h}$. For the gas-liquid systems the susceptibility (compressibility) determines how fast the averaged density of the system changes with the change of the external pressure: $K \propto \frac{\partial \rho}{\partial P}$. Following other authors [13, 15, 16], for a system with damage we could define the susceptibility as a quantity that determines how fast the order parameter $D$ changes with the change of the external stress $\sigma$: $\chi \propto \frac{\partial D_0}{\partial \sigma}$. For the vicinity of the spinodal point from Eq. (5) we immediately obtain [13] that the exponent $\gamma$ of this susceptibility equals to 1/2: $\chi \propto |\sigma - \sigma_S|^{-\gamma} \propto 1/\sqrt{|\sigma - \sigma_S|}$.

So defined susceptibility indeed corresponds to the rate of change of the order parameter $D$ with the change of the field parameter $\sigma$. However, we should remember that the choice of the field parameter is not unique, especially for our case of the presence of geometric constraint (1). For example, we could choose the strain $E\varepsilon = \sigma / (1 - D)$ to be a field parameter. Or we could choose an arbitrary function of $\sigma$. Therefore, to identify the true susceptibility we need to check that this quantity will satisfy the fluctuation-dissipation theorem.

Unfortunately, the statistical mechanics of our system under the BC $\sigma$ = const is non-Gibbsian [12]. Therefore, to understand what the fluctuations in the model are we have to start from the simpler BC $\varepsilon$ = const. For this BC it was shown in [12] that the behavior of the system is analogous to the behavior of the canonical ensemble.

We imagine now the canonical ensemble of a simple spin system, e.g. Ising model, whose Hamiltonian is $H_{\{\eta\}} = -J\sum_{<i,j>} R_{ij}\eta_i\eta_j - h\sum_{i=1}^{N}\eta_i$ where index '$\{\eta\}$' indicates a particular microstate (a particular realization of spins on the lattice). Magnetization of this microstate is $Nm_{\{\eta\}} \equiv \sum_{i=1}^{N}\eta_i$. The partition function of this system is $Z \equiv \sum_{\{\eta\}} e^{-H_{\{\eta\}}/T}$. The partial derivative of the partition function $Z$ over the magnetic field gives $\frac{\partial Z}{\partial h} = \sum_{\{\eta\}}\left(\frac{1}{T}e^{-H_{\{\eta\}}/T}\sum_{i=1}^{N}\eta_i\right) = Z\frac{N\langle m_{\{\eta\}}\rangle^{equil}}{T}$. So, the averaged magnetization in the equilibrium is $N\langle m_{\{\eta\}}\rangle^{equil} = T\frac{\partial \ln Z}{\partial h}$. In the similar way we obtain $\langle (Nm_{\{\eta\}})^2\rangle^{equil} = \frac{T^2}{Z}\frac{\partial^2 Z}{\partial h^2}$ and $\langle (Nm_{\{\eta\}} - N\langle m_{\{\eta\}}\rangle^{equil})^2\rangle^{equil} = T^2\frac{\partial^2 \ln Z}{\partial h^2}$. Therefore for the susceptibility we have

$$\chi \propto \frac{\partial \langle Nm_{\{\eta\}} \rangle^{equil}}{\partial h} = T\frac{\partial^2 \ln Z}{\partial h^2} = \frac{1}{T}\left\langle \left(Nm_{\{\eta\}} - \langle Nm_{\{\eta\}} \rangle^{equil}\right)^2 \right\rangle^{equil}.$$ This is the fluctuation-dissipation theorem: we have made a connection between the sensitivity of the system order parameter relative to the field parameter and the variance of fluctuations of this order parameter.

For the FBM we can follow the same way of conclusions. As we have shown in [12] for the BC $\varepsilon$ = const we can rewrite the probability of microstates (2a) as $w_{\{n\}}^{equil}(D) = \frac{1}{Z}e^{-ND/T}$ where $Z = (1 - P_s(E\varepsilon))^{-N}$ is the partition function of the system $Z = \sum_{\{n\}} e^{-ND/T}$. The role of the topological temperature $T$ is played by the quantity $T = \ln^{-1}\frac{1 - P_s(E\varepsilon)}{P_s(E\varepsilon)}$. In the way similar to the canonical ensemble of the Ising model we obtain $N\langle D \rangle^{equil} = -\frac{\partial \ln Z}{\partial (1/T)}$, $\langle (ND)^2 \rangle^{equil} = \frac{1}{Z}\frac{\partial^2 Z}{\partial (1/T)^2}$, and $\left\langle \left(ND - N\langle D \rangle^{equil}\right)^2 \right\rangle^{equil} = \frac{\partial^2 \ln Z}{\partial (1/T)^2}$. To make a connection with the fluctuation-dissipation theorem for the susceptibility we need to assume $\chi \propto \left\langle \left(ND - N\langle D \rangle^{equil}\right)^2 \right\rangle^{equil} = \frac{\partial^2 \ln Z}{\partial (1/T)^2} = -\frac{\partial N\langle D \rangle^{equil}}{\partial (1/T)}$. Remembering that the temperature $T$ is the function of $P_s(E\varepsilon)$ and also that for the equilibrium we have $N\langle D \rangle^{equil} = ND_0 = NP_s(E\varepsilon)$ we obtain

$$\chi \propto ND_0(1 - D_0). \tag{8}$$

In the vicinity of spinodal $\chi \propto ND_{0S}(1 - D_{0S})$. Therefore the spinodal exponent $\gamma$ of the susceptibility for the BC $\varepsilon$ = const equals 0. We could expect this result because for this BC the fibers do not interact and for the prescribed fiber's strain there is nothing special about the fluctuations in the vicinity of the spinodal point.

For the BC $\sigma$ = const we cannot follow the same way of deduction because the behavior of the system in this case is non-Gibbsian [12]. However, we can discover (Eq. (8) of reference [12]) that for the BC $\varepsilon$ = const the right-hand side of the Eq. (8) is the squared width of the maximum of (4a). This should be expected because the function given by Eq. (4a) is the probability to find a macrostate [D] in equilibrium. The width of the maximum of this function determines the standard deviation of Gaussian fluctuations in the vicinity of the equilibrium:

$$\ln W_{[D]}^{equil}(D) = \ln W_{[D]}^{equil}(D_0) + \frac{1}{2}\frac{d^2 \ln W_{[D]}^{equil}}{dD^2}(D_0) \cdot (D-D_0)^2 \text{ or } W_{[D]}^{equil}(D) \propto \exp\left(-\frac{N^2(D-D_0)^2}{2ND_0(1-D_0)}\right) \text{ for}$$

the BC $\varepsilon$ = const. For the BC $\sigma$ = const we respectively obtain

$$\ln W_{[D]}^{equil}(D) = \ln W_{[D]}^{equil}(D_0) - \frac{1}{2}\frac{1}{ND_0(1-D_0)}\left\{1 - \frac{\sigma}{(1-D_0)^2}P_s'\left(\frac{\sigma}{1-D_0}\right)\right\}^2 \cdot N^2(D-D_0)^2 + ... \text{ For the}$$

equilibrium in this case we have $D_0 = P_s\left(\frac{\sigma}{1-D_0}\right)$ and the differential of this equation given by Eq. (6). Then for the fluctuations $(D-D_0)$ around the maximum of $W_{[D]}^{equil}(D)$ we obtain $W_{[D]}^{equil}(D) \propto \exp\left\{-\frac{N^2(D-D_0)^2}{2ND_0(1-D_0)\left(1+\frac{\sigma}{1-D_0}\frac{dD_0}{d\sigma}\right)^2}\right\}$ where $\frac{dD_0}{d\sigma}$ is the derivative corresponding to the change of the equilibrium damage $D_0$ with the equilibrium change of the constant external stress $\sigma$. This gives us an opportunity to assume that the

susceptibility for the BC $\sigma$ = const, as a measure of fluctuations, can be hypothesized to be

$$\chi \propto ND_0(1-D_0)\left\{1-\frac{\sigma}{(1-D_0)^2}P'_s\left(\frac{\sigma}{1-D_0}\right)\right\}^{-2} = ND_0(1-D_0)\left(1+\frac{\sigma}{1-D_0}\cdot\frac{dD_0}{d\sigma}\right)^2. \tag{9}$$

This is the representation of the susceptibility by fluctuations. Unfortunately, we do not know how to reinstate the fluctuation-dissipation theorem in this case, *i.e.* how to associate this expression with the derivative of the order parameter by some field parameter.

In the vicinity of the spinodal we have

$$\frac{d\sigma}{dD_0}(D_0) = \left.\frac{d\sigma}{dD_0}\right|_S + \left.\frac{d\frac{d\sigma}{dD_0}}{dD_0}\right|_S (D_0 - D_{0S}) = \left.\frac{d^2\sigma}{dD_0^2}\right|_S (D_0 - D_{0S}) \text{ and}$$

$$\chi \propto N\frac{D_S}{1-D_S}\frac{\sigma_S^2}{\left(\left.\frac{d^2\sigma}{dD_0^2}\right|_S\right)^2}|\sigma_0 - \sigma_{0S}|^{-1} \tag{10}$$

and therefore the spinodal exponent $\gamma$ of the susceptibility for the BC $\sigma$ = const equals 1. Here we again assume that $P_s(s)$ does not have any singularity at the spinodal point and therefore the second derivative $\left.\frac{d^2\sigma}{dD_0^2}\right|_S$ does not equal to zero.

For the canonical ensemble of an arbitrary thermodynamic system with Hamiltonian $H_{\{\eta\}}$ the partition function is $Z \equiv \sum_{\{\eta\}} e^{-H_{\{\eta\}}/T}$. The partial derivative of the partition function $Z$ over the quantity $1/T$ gives $\frac{\partial Z}{\partial(1/T)} = -\sum_{\{\eta\}}\left(e^{-H_{\{\eta\}}/T}H_{\{\eta\}}\right) = -Z\langle H_{\{\eta\}}\rangle^{equil}$.

Therefore the averaged energy in the equilibrium is $\langle H_{\{\eta\}} \rangle^{equil} = -\frac{\partial \ln Z}{\partial (1/T)}$. In the similar way we get $\langle H_{\{\eta\}}^2 \rangle^{equil} = \frac{1}{Z} \frac{\partial^2 Z}{\partial (1/T)^2}$ and $\langle (H_{\{\eta\}} - \langle H_{\{\eta\}} \rangle^{equil})^2 \rangle^{equil} = \frac{\partial^2 \ln Z}{\partial (1/T)^2}$. We see here the direct analogy with the formulas we used above to find the susceptibility for the BC $\varepsilon$ = const. But in statistical mechanics the specific heat $C$ is proportional to $\frac{\partial^2 \ln Z}{\partial (1/T)^2}$ and, correspondently, to the variance of energy in the vicinity of the equilibrium: $\langle (H_{\{\eta\}} - \langle H_{\{\eta\}} \rangle^{equil})^2 \rangle^{equil}$. We were looking for the susceptibility as a variance of the order parameter but what we found was actually the specific heat. This result is expected. Indeed, the geometric constraint (1), which acts even on the level of microstates, makes for our model one field parameter less. As it was obtained in [12], the analogue of the energy balance equation is the equation of topological balance $NdD_0 = Td\langle S \rangle^{equil}$. From this equation we see that the role of the energy in the canonical ensemble is played by the order parameter $D_0$ and on the right-hand side of the equation we have only a topological analogue of the heat in the canonical ensemble. Therefore for our system the specific heat, as the variance of the energy analogue, is proportional to the susceptibility, as the variance of the order parameter. Therefore for all BCs we expect that the spinodal exponent $\alpha$ for the specific heat should coincide with the spinodal exponent $\gamma$ for the susceptibility.

For the BC $\sigma$ = const we found the expression for the susceptibility (9) by the fluctuation part of the fluctuation-dissipation theorem. But we still do not know how this susceptibility can be associated with the dissipation in the system. In other words, we can only guess what field parameter $X$ we should use for the derivative of the order parameter $D_0$ to obtain required Eq. (9): $\chi \propto \dfrac{dD_0}{dX}$ where

$$X \propto \int \frac{dD_0}{ND_0(1-D_0)}\left\{1-\frac{\sigma}{(1-D_0)^2}P'_s\left(\frac{\sigma}{1-D_0}\right)\right\}^2 = \int \frac{dD_0}{ND_0(1-D_0)}\left(1+\frac{\sigma}{1-D_0}\cdot\frac{dD_0}{d\sigma}\right)^{-2}.$$

## 6 The role and scaling of free energy potential

The main purpose of this section is to find the scaling behavior of the free energy potential of a metastable state. Therefore first we need to discuss what the metastable state is and how we can find its free energy. Again we should start from a simple example, e.g. Ising model, and discuss the role that the free energy potential plays in statistical mechanics.

We consider the canonical ensemble of a meanfield Ising model whose Hamiltonian $H_{\{\eta\}}$ of microstate $\{\eta\}$ depends only on the magnetization $m_{\{\eta\}} \equiv \dfrac{1}{N}\sum_{i=1}^{N}\eta_i$ of this microstate: $H_{\{\eta\}} = H_{\{\eta\}}(m_{\{\eta\}})$. As to a microstate $\{\eta\}$ we refer to a particular realization of spin orientations on the lattice. As to a macrostate $[m]$ we will refer to the union of all microstates corresponding to the given value of $m$: $[m] = \bigcup_{\{\eta\}:m_{\{\eta\}}=m}\{\eta\}$. Following [21], as to a metastable state !*local*! we will refer to the union of all microstates or macrostates in the vicinity of the local minimum $m_0^{local}$ of the free energy potential: $!local! = \bigcup_{\{\eta\}\in \text{vicinity of } m_0^{local}}\{\eta\} = \bigcup_{[m]\in \text{vicinity of } m_0^{local}}[m]$. As to a globally stable state !*global*! we will refer to the union of all microstates or macrostates in the vicinity of the global minimum $m_0^{global}$ of the free energy potential: $!global! = \bigcup_{\{\eta\}\in \text{vicinity of } m_0^{global}}\{\eta\} = \bigcup_{[m]\in \text{vicinity of } m_0^{global}}[m]$. As to the equilibrium state of the system in equilibrium with the prescribed temperature of the canonical ensemble we will refer to the union of all microstates or all macrostates:

$$!equil! = \bigcup_{\{\eta\}} \{\eta\} = \bigcup_{[m]} [m].$$

The equilibrium probability of a microstate is the function only of its magnetization: $w_{\{\eta\}}^{equil} = \frac{1}{Z} e^{-H_{\{\eta\}}/T} \equiv w^{equil}(m_{\{\eta\}}) = \frac{1}{Z} e^{-H(m_{\{\eta\}})/T}$. For each macrostate $[m]$ we will denote the number of corresponding microstates as $g_{[m]}$. For the Ising model $g_{[m]} = \frac{N!}{(N(1+m)/2)!(N(1-m)/2)!}$ however further we will use arbitrary dependence of $g_{[m]}$ on $m$ assuming only its fast exponential growth.

The probability of a macrostate $[m]$ to be observed in equilibrium with the externally prescribed temperature is $W_{[m]}^{equil} = g_{[m]} w^{equil}(m)$. Both functions $g_{[m]}$ and $w^{equil}(m)$ have exponential dependence on $m$ where the exponent is proportional to the number of spins $N$. Therefore in the thermodynamic limit the dependence $W_{[m]}^{equil}$ on $m$ has very sharp maxima at the stable value $m = m_0^{global}$ and at the metastable value $m = m_0^{local}$. Therefore, the probability of the metastable state $!local!$ as the sum of probabilities of macrostates in the vicinity of $m = m_0^{local}$ is equal to the probability of the macrostate $[m_0^{local}]$: $W_{!local!}^{equil} = \sum_{\{\eta\} \in \text{vicinity of } m_0^{local}} w_{\{\eta\}}^{equil} = \sum_{[m] \in \text{vicinity of } m_0^{local}} W_{[m]}^{equil} \approx_{\ln} g_{[m_0^{local}]} w^{equil}(m_0^{local}) = W_{[m_0^{local}]}^{equil}$. The same is valid for the global maximum: $W_{!global!}^{equil} = \sum_{\{\eta\} \in \text{vicinity of } m_0^{global}} w_{\{\eta\}}^{equil} = \sum_{[m] \in \text{vicinity of } m_0^{global}} W_{[m]}^{equil} \approx_{\ln} g_{[m_0^{global}]} w^{equil}(m_0^{local}) = W_{[m_0^{global}]}^{equil}$. Of course, $W_{!local!}^{equil} + W_{!global!}^{equil} \approx 1$. The probability of the equilibrium state $!equil!$ as the sum of probabilities of all microstates in equilibrium with the prescribed temperature equals, of course, unity: $W_{!equil!}^{equil} = \sum_{\{\eta\}} w_{\{\eta\}}^{equil} = \sum_{[m]} W_{[m]}^{equil} = 1$.

Let us introduce partial partition functions. For a microstate $\{\eta\}$ we define the partial partition function $Z_{\{\eta\}}$ as $Z_{\{\eta\}} \equiv e^{-H_{\{\eta\}}/T} = Z w_{\{\eta\}}^{equil}$, i.e. only a single exponential factor of this macrostate. For a macrostate $[m]$ we define the partial partition function $Z_{[m]}$ as $Z_{[m]} \equiv \sum_{\{\eta\} \in [m]} e^{-H_{\{\eta\}}/T} = Z W_{[m]}^{equil}$, i.e. the sum of exponential factors only over the microstates corresponding to the given macrostate. For the metastable state !local! we define the partial partition function $Z_{!local!}$ as $Z_{!local!} \equiv \sum_{\{\eta\} \in \text{vicinity of } m_0^{local}} e^{-H_{\{\eta\}}/T} \equiv \sum_{[m] \in \text{vicinity of } m_0^{local}} g_{[m]} e^{-H(m)/T} = Z W_{!local!}^{equil}$. The same we can do for the global minimum: $Z_{!global!} \equiv \sum_{\{\eta\} \in \text{vicinity of } m_0^{global}} e^{-H_{\{\eta\}}/T} \equiv \sum_{[m] \in \text{vicinity of } m_0^{global}} g_{[m]} e^{-H(m)/T} = Z W_{!global!}^{equil}$. For the equilibrium state !equil! the partial partition function is the total partition function of the system: $Z_{!equil!} = \sum_{\{\eta\}} e^{-H_{\{\eta\}}/T} = \sum_{[m]} g_{[m]} e^{-H(m)/T} \equiv Z$. Of course, $Z_{!local!} + Z_{!global!} \approx_{\ln} Z$.

We have obtained that the probabilities of a microstate, macrostate, metastable state, stable state, and equilibrium state are the relative partial partition functions of these states:

$w_{\{\eta\}}^{equil} = Z_{\{\eta\}}/Z$, $W_{[m]}^{equil} = Z_{[m]}/Z$, $W_{!local!}^{equil} = Z_{!local!}/Z$, $W_{!global!}^{equil} = Z_{!global!}/Z$, and $W_{!equil!}^{equil} = Z_{!equil!}/Z$ .(11)

Next step is to introduce the entropy of these states. We will follow [12]. First we will imagine a system isolated on one microstate $\{\eta'\}$. Only one microstate corresponds to this microstate $\{\eta'\}$. For the system isolated on this microstate the probability of this microstate is, of course, unity: $w_{\{\eta'\}} = 1$. The entropy of the system isolated on this microstate is $S_{\{\eta'\}} \equiv -\sum_{\{\eta\}:\{\eta\}=\{\eta'\}} w_{\{\eta\}} \ln w_{\{\eta\}} = -w_{\{\eta'\}} \ln w_{\{\eta'\}} = 0$. We did not use here the superscript 'equil' for the probabilities and entropy because the system isolated on a microstate is not in equilibrium with the prescribed temperature of the canonical ensemble. In other words, isolation of the system on a microstate corresponds to a non-equilibrium state for the canonical ensemble.

We imagine now a system isolated on a macrostate $[m]$. $g_{[m]}$ microstates correspond to this macrostate. Probabilities of these microstates for the system isolated on the given macrostate $[m]$ are $w_{\{\eta\}} = 1/g_{[m]}$. Therefore the entropy of the system isolated on this macrostate is $S_{[m]} \equiv -\sum_{\{\eta\}\in[m]} w_{\{\eta\}} \ln w_{\{\eta\}} = -\frac{g_{[m]}}{g_{[m]}} \ln \frac{1}{g_{[m]}} = \ln g_{[m]}$.

Now we need to find the entropy of the system isolated on the metastable state !*local*!. However now we need to be more specific than [21] and restrict the definition what we do understand under the vicinity of the local minimum of the free energy potential. In other words, how many microstates or macrostates we should include in the vicinity of the point $m_0^{local}$. Natural way is to restrict the metastable state to the width of the local maximum of $W_{[m]}^{equil} = g_{[m]} w^{equil}(m)$ which determines the size of fluctuations. The number of different macrostates $[m]$ in the width of the local maximum of $W_{[m]}^{equil}$ has the power-law dependence on $N$ while the number of microstates corresponding to each of this macrostates $[m]$ is $g_{[m]}$ and has the exponential dependence on $N$. Therefore the number of microstates corresponding to the metastable state (which are in the width of the maximum of $W_{[m]}^{equil}$) with the logarithmical accuracy equals to the number of microstates corresponding to one of the macrostates in the maximum: $g_{!local!} \approx_{\ln} g_{[m_0^{local}]}$. Because the local maximum of $W_{[m]}^{equil}$ is very sharp, the probabilities of these microstates with the logarithmical accuracy equal to $w_{\{\eta\}} \approx_{\ln} 1/g_{!local!} \approx_{\ln} 1/g_{[m_0^{local}]}$. Therefore for the system isolated on the metastable state !*local*! the entropy equals the entropy of the macrostate $[m_0^{local}]$:

$$S_{!local!} \equiv -\sum_{\{\eta\}\in!m_0^{local}!} w_{\{\eta\}} \ln w_{\{\eta\}} = -\frac{g_{!local!}}{g_{!local!}} \ln \frac{1}{g_{!local!}} = \ln g_{!local!} \approx \ln g_{[m_0^{local}]} = S_{[m_0^{local}]}.$$ In the similar way we

obtain $Z_{!local!} \equiv \sum_{[m]\in \text{ vicinity of } m_0^{local}} g_{[m]} e^{-H_{\{\eta\}}(m)/T} \approx_{\ln} g_{[m_0^{local}]} e^{-H_{\{\eta\}}(m_0^{local})/T} = Z_{[m_0^{local}]}$. The same is valid and for the global minimum of the free energy potential: $S_{!global!} = \ln g_{!global!} \approx \ln g_{[m_0^{global}]} = S_{[m_0^{global}]}$ and

$Z_{!global!} \equiv \sum_{[m]\in \text{ vicinity of } m_0^{global}} g_{[m]} e^{-H_{\{\eta\}}(m)/T} \approx_{\ln} g_{[m_0^{global}]} e^{-H_{\{\eta\}}(m_0^{global})/T} = Z_{[m_0^{global}]}$. For the equilibrium state

!*equil*! the system isolated on this state is our system in equilibrium with the canonical ensemble. Therefore the probabilities of microstates equal to the probabilities in equilibrium with the prescribed temperature: $w_{\{\eta\}} = w_{\{\eta\}}^{equil}$. Therefore for the entropy of this state we have $S_{!equil!} \equiv -\sum_{\{\eta\}\in!equil!} w_{\{\eta\}} \ln w_{\{\eta\}} = -\sum_{\{\eta\}} w_{\{\eta\}}^{equil} \ln w_{\{\eta\}}^{equil}$. Remembering, that both maxima of $W_{[m]}^{equil}$ are very sharp we obtain

$S_{!equil!} \approx_{\ln} -g_{!local!} w^{equil}(m_0^{local}) \ln w^{equil}(m_0^{local}) - g_{!global!} w^{equil}(m_0^{global}) \ln w^{equil}(m_0^{global})$ or

$S_{!equil!} \approx_{\ln} -W_{!local!}^{equil} \ln w^{equil}(m_0^{local}) - W_{!global!}^{equil} \ln w^{equil}(m_0^{global})$.

We define the Helmholtz free energy as $A \equiv E - TS = \sum_{\{\eta\}} w_{\{\eta\}}(H_{\{\eta\}} + T \ln w_{\{\eta\}})$ for equilibrium and non-equilibrium states where $E$ is the averaged energy of this state. For a system isolated on a microstate $\{\eta\}$ we have $A_{\{\eta\}} \equiv H_{\{\eta\}} - TS_{\{\eta\}} = -T \ln Z_{\{\eta\}}$. For a system isolated on a macrostate $[m]$ the probabilities of the corresponding microstates are $w_{\{\eta\}} = 1/g_{[m]}$ and we have $A_{[m]} \equiv \sum_{\{\eta\}\in[m]} w_{\{\eta\}} H_{\{\eta\}} - TS_{[m]} = \frac{g_{[m]}}{g_{[m]}} H(m) - TS_{[m]} = -T \ln Z_{[m]}$. For a system isolated on the metastable state $!local!$ we have $w_{\{\eta\}} \approx_{\ln} 1/g_{!local!} \approx_{\ln} 1/g_{[m_0^{local}]}$ and

$A_{!local!} \equiv \sum_{\{\eta\}\in !local!} w_{\{\eta\}} H_{\{\eta\}} - TS_{!local!} = \frac{g_{!local!}}{g_{!local!}} H(m_0^{local}) - TS_{!local!} = -T \ln Z_{!local!}$. For a system isolated on the stable state $!global!$ we have $A_{!global!} = -T \ln Z_{!global!}$. For the system of the equilibrium state $!equil!$ we have $w_{\{\eta\}} = w_{\{\eta\}}^{equil}$ and

$A_{!equil!} \equiv \sum_{\{\eta\}\in !equil!} w_{\{\eta\}} H_{\{\eta\}} - TS_{!equil!} = g_{!local!} w^{equil}(m_0^{local}) H(m_0^{local}) + g_{!global!} w^{equil}(m_0^{global}) H(m_0^{global}) - TS_{!equil!}$.

Remembering formula for the $S_{!equil!}$ we obtain

$A_{!equil!} = W_{!local!}^{equil} \{H(m_0^{local}) + T \ln w^{equil}(m_0^{local})\} + W_{!global!}^{equil} \{H(m_0^{global}) + T \ln w^{equil}(m_0^{global})\}$ or

$A_{!equil!} = -T \ln Z \{W_{!local!}^{equil} + W_{!global!}^{equil}\} \approx -T \ln Z = -T \ln Z_{!equil!}$. Therefore the Helmholtz free energy, which is the free energy potential for the canonical ensemble, for any non-equilibrium state equals minus logarithm of the partial partition function of this state times temperature.

For the probabilities (11) of a macrostate, macrostate, and metastable state respectively in equilibrium with the prescribed temperature as the BC of the canonical ensemble we obtain

$$w_{\{\eta\}}^{equil} = \frac{1}{Z} e^{-A_{\{\eta\}}/T}, \tag{12a}$$

[21, 22]: $W_{[m]}^{equil} = \frac{1}{Z} e^{-A_{[m]}/T}$, $W_{!local!}^{equil} = \frac{1}{Z} e^{-A_{!local!}/T}$, and $W_{!global!}^{equil} = \frac{1}{Z} e^{-A_{!global!}/T}$. (12b)

Therefore when somebody reads in a textbook that the distribution of probabilities of microstates in the canonical ensemble is Boltzmann's distribution of energies, in fact, it should be read as Boltzmann's distribution not of energies but of free energy potentials. Only due to the fact that the entropy of one microstate is zero it becomes Boltzmann's distribution of energies.

There are some differences between the considered canonical ensemble of the Ising model and the FBM. For the BC $\varepsilon = $ const there is only one maximum of $W_{[D]}^{equil}$ for any specified value of strain. For the state !A! (Fig. 1) we have $A_{!A!} = -T \ln Z_{!A!} \approx -T \ln Z_{[D_{0A}]}$. Remembering (11), $A_{!A!} \approx -T \ln ZW_{[D_{0A}]}^{equil} = -T \ln g_{[D_{0A}]} e^{-ND_{0A}/T}$. We see, that this expression is not singular when the state !A! approaches the spinodal point S. Therefore we have confirmed that for the susceptibility and specific heat, which are proportional to $\dfrac{d^2 A}{d(1/T)^2}$, for the BC $\varepsilon = $ const we have zero spinodal exponents.

For the BC $\sigma = $ const we in general do not know the free energy potential. Indeed, for this BC the behavior of the model is non-Gibbsian [12]. The probabilities of microstates with damage $D$ are

$$w^{equil}(D) = \exp\left( N(1-D) \ln\left[ 1 - P_s\left( \frac{\sigma}{1-D} \right) \right] + ND \ln\left[ P_s\left( \frac{\sigma}{1-D} \right) \right] \right) \qquad (13)$$

and we cannot reconstruct them in such a way that the linear function of the order parameter would be in the exponent. We could try to linearize probabilities (13) in the vicinity of the metastable state !A!. The width of the local maximum A of $W_{[D]}^{equil}$ is proportional to $1/\sqrt{N}$ far from the spinodal and to $1/\sqrt[4]{N}$ in the vicinity of the spinodal. Therefore, around the quasi-equilibrium metastable state A, the range of fluctuations with non-zero probabilities interesting for us is also proportional to $1/\sqrt{N}$ or $1/\sqrt[4]{N}$ and in the thermodynamic limit is very narrow. We could try to use Taylor expansion of the exponent of (13) in the vicinity of the metastable state

$$w^{equil}(D) = \exp\left(N(1-D_{0A})\ln(1-D_{0A}) + ND_{0A}\ln D_{0A} - N(D-D_{0A})/T_A\right) \text{ or} \quad (14a)$$

$$w^{equil}(D) \approx_{\ln} \frac{1}{g_{[D_{0A}]}} e^{-\frac{N(D-D_{0A})}{T_A}} = \frac{1}{g_{[D_{0A}]} e^{-\frac{ND_{0A}}{T_A}}} e^{-\frac{ND}{T_A}} \equiv \frac{1}{Z_A} e^{-\frac{ND}{T_A}} \quad (14b)$$

where $D_{0A}$ is the equilibrium value of damage in the metastable state and $T_A \equiv \ln^{-1}\frac{1-D_{0A}}{D_{0A}}$ is the virtual temperature. $Z_A \equiv g_{[D_{0A}]} e^{-\frac{ND_{0A}}{T_A}}$ plays here the role of the partition function as the sum of factors $e^{-\frac{ND}{T_A}}$ in the vicinity of the metastable state.

However, this approach does not work. Although the maximum of the function $W^{equil}_{[D]}$ is very narrow both functions $g_{[D]}$ and $w^{equil}$ have $N$ in their exponents. Therefore their change in the width of the maximum is very rapid and we cannot use the linear approximation for their exponents. This linearization would significantly change the fluctuating behavior. This can be demonstrated by finding the susceptibility - specific heat $\chi \propto \left\langle \left(ND - N\langle D\rangle_A^{equil}\right)^2\right\rangle_A^{equil} = \frac{\partial^2 \ln Z_A}{\partial(1/T_A)^2} = -T^3\frac{d^2 A_A^{equil}}{dT_A^2} = -\frac{\partial N\langle D\rangle_A^{equil}}{\partial(1/T_A)}$ where $A_A \equiv ND - T_A S$ for equilibrium and non-equilibrium states and for the metastable equilibrium $A_A^{equil} = -T_A \ln Z_A$. The behavior of the susceptibility is significantly different from found in Section 5 and non-Gibbsian mechanics, as it could be expected, appears to be more complex than its projection onto the Gibbsian analogue. Therefore we cannot construct an analogy with the canonical ensemble for the metastable state of this system. The analogy with the Boltzmann distribution is no more valid, therefore we cannot define the temperature and free energy potential for the BC $\sigma$ = const.

**7 Conclusions**

For the spinodal point of the model with damage we find scaling behavior similar to the spinodal scaling behavior in statistical mechanics. The susceptibility of the model is shown to be more complex quantity than the partial derivative of the damage variable by the external stress. In fact, the susceptibility is shown to be proportional to the specific heat. Also we find the analytical expression for the free energy potential. For the external boundary constraint $\varepsilon$ = const the spinodal exponents are found to be $\alpha = 0$, $\beta = 1/2$, $\gamma = 0$, and $\tau = 0$. For the external boundary constraint $\sigma$ = const the spinodal exponents are found to be $\alpha = 1$, $\beta = 1/2$, $\gamma = 1$, and $\tau = 1$. While our results have more general applicability than only for the meanfield fiber-bundle model, used as an example, the found exponents clearly have typical meanfield rational values. The exponent equalities, *e.g.* $\alpha + 2\beta + \gamma = 2$, are not satisfied but it would be difficult to expect their validity for the degenerated case of only one field parameter when the specific heat coincides with the susceptibility. However, for the external boundary constraint $\sigma$ = const we still see their representation: *e.g.,* $\alpha + 2\beta = 2$ or $2\beta + \gamma = 2$.

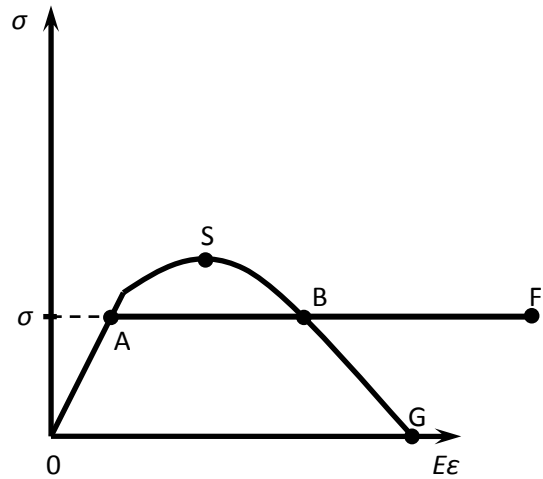

Fig. 1. Phase diagram. Point S is the spinodal point of the model. Curve 0 – A – S represents the metastable states. Point F is the point of complete fracture and is assumed to represent infinite strain. Straight line A - B - F corresponds to Maxwell's rule.